\documentclass[pra,twocolumn,aps]{revtex4}
\usepackage{graphics}

\begin{document}

\title{Ray-Tracing studies in a perturbed atmosphere: I- The initial value problem.}
\author{C. Tannous}
\altaffiliation{Present address: Laboratoire de Magnétisme de Bretagne, UPRES A CNRS 6135,
Université de Bretagne Occidentale, BP: 809 Brest CEDEX, 29285 FRANCE}
\affiliation{Alberta Government Telephones, Calgary, CANADA T2G 4Y5}
\author{J. Nigrin}
\affiliation{Alberta Government Telephones,Floor 5, 
10065 Jasper Av., Edmonton, CANADA T5J 3B1}

\date{March 28, 2001}

\begin{abstract}
We  report  the development of a new ray-tracing  simulation
tool having the potential of the full characterization of  a
radio  link  through the accurate study of  the  propagation
path  of  the signal from the transmitting to the  receiving
antennas  across  a  perturbed atmosphere.  The  ray-tracing
equations  are  solved, with controlled accuracy,  in  three
dimensions  (3D)  and  the propagation  characteristics  are
obtained   using  various  refractive  index   models.   The
launching  of  the  rays,  the atmospheric  medium  and  its
disturbances  are characterized in 3D. The  novelty  in  the
approach  stems from the use of special numerical techniques
dealing  with so called stiff differential equations without
which  no solution of the ray-tracing equations is possible.
Starting with a given launching angle, the solution consists
of  the ray trajectory, the propagation time information  at
each  point of the path, the beam spreading, the transmitted
(resp.  received)  power  taking account  of  the  radiation
pattern  and  orientation of the antennas and  finally,  the
polarization  state  of  the  beam.  Some  previously  known
results  are  presented  for comparative  purposes  and  new
results are presented as well as some of the capabilities of
the software.


\end{abstract}

\maketitle

\section{Introduction}

Multipath propagation is believed to be the major  cause  of
data  transmission impairments in terrestrial line of  sight
microwave radio systems.
Efficient antenna design requires the understanding  of  the
propagation  of  individual  rays  across  the  channel  and
gauging  the  refractive  index of the  various  atmospheric
disturbances    any   given  ray   encounters   during   its
propagation. Adopting a refractive index model for  a  given
disturbance  arising from spatial fluctuations in  humidity,
pressure   or  temperature  (these  fluctuations  might   be
temporal as well, but we shall consider, for the time being,
that  the  propagation  time occurs on  a  time  scale  much
smaller  than  the one associated with these  fluctuations),
we  establish the ray propagation equations and  solve  them
with several numerical techniques having a first, fourth and
sixth   order  accuracy.  The  ray  tracing  equations   are
initially solved in two dimensions bypassing the effects  of
small  and non-linear terms as explained in section 2. Later
on, we switch to 3D in order to assess the effects the small
and  non-linear terms have on ray propagation. Several facts
emerge from this approach:
\begin{itemize}
\item The  small  non-linear terms  lead  to  a  breakdown  of
standard  integration techniques. The  ray  equations  which
constitute   a  system  of  6  ordinary  coupled  non-linear
differential   equations  become  stiff.  This   means   the
integration  step becomes so small (because of the  presence
of  terms that differ by several orders of magnitude) making
the integration process so slow that any progress in seeking
a solution of the system is virtually stopped.
\item The relation between the launching and arrival angles for
a   given  disturbance  are  profoundly  altered.  What  was
previously believed to be a "good" or "bad" launching  angle
might have gotten its true attributes from reasons different
from what is currently known.
\item  A  very  high  sensitivity is  observed  around  certain
launching  angles: a very small uncertainty in the launching
angle  can induce the ray to take a path radically different
from what is normally expected.
\end{itemize}

This report is organized in the following way: In section 2,
we  establish the ray-tracing equations (RTE). In section  3
we  describe  some  of the problems encountered  during  the
solution  of the RTE, namely those related to stiffness  and
present  the algorithms to cure them (Appendix A contains  a
description and an example of a stiff system). In section  4
we  compare  our approach to previous ones and present  some
illustrative  new  cases in section  5.  This  section  also
describes the potential applications of the software and its
capabilities.  Section  6 discusses some  possibilities  for
future  developments. Appendix B shows how  to  avoid  stiff
differential  equations in two dimensions and turn  the  RTE
into a set of recursion relations.

\section{Ray tracing equations}

In   terrestrial  microwave  radio  systems,  the  range  of
frequencies  used  and in comparison  the  range  of  length
scales  present in the channel allow us to use  a  geometric
(or   ray)  approach  to  electromagnetic  propagation.  The
fundamental  equation of geometrical optics is  the  Eikonal
equation :

\begin{equation}
{({\bf grad} S)}^2 = n^2
\end{equation}

where n is the local refractive index and S is the local
phase of the ray.  Taking the gradient of both sides of the
Eikonal equation gives  the second order vector propagation
equation:
\begin{equation}
\frac{ d (\frac{n d{\bf R}}{ds} )}  {ds} = {\bf grad} n
\end{equation}
                              
where  ${\bf R}$  is  the  ray  position and $ds$  is  a  differential
displacement along the ray path, i.e. $ds=||d{\bf R}||$,  the norm of
the vector $d{\bf R}$.

This can be rewritten as a system of two first order
equations:

\begin{eqnarray}
\frac{d {\bf R}}{ds} &=& {\bf T}        \nonumber \\
\frac{d(n {\bf T})} {ds}    &=& \mbox{{\bf grad}} n
\end{eqnarray}

where  ${\bf T}$  is  a  unit vector tangent to the  ray  path  (The
geometry  is depicted in Fig.1). The advantage of solving  a
first  order system rather than a single second order system
is threefold:\\

\begin{itemize}
\item Stability problems are easier to handle.
\item Validity of the solution is easy to monitor since one  has
to  have  for  all  times $||{\bf T}||=1$  
providing  a  simple  means  to check  the  quality  of  the
integration procedure.
\item Accuracy  of  the  solution is controlled  within  certain
tolerance limits depending on the selected integration step.
\end{itemize}

This is discussed in detail in section 5.
The  refractive index function of the atmosphere is  written
as:

\begin{equation}
 n=1+10^{-6} N
\end{equation}

where  $N$  depends on the frequency used, humidity conditions
and  height above the Earth ground. Several models exist for
the range of frequencies and heights we are dealing with and
are generally expressed in N units. The following two models
are  of interest; the first for a normal atmosphere and  the
second for a disturbed one:\\

\begin{itemize}
\item Exponential model: $N=315 \mbox{ exp}(-0.136 h)$, with h (height) in kms.   
\item Webster model:    $N=300.+kh+ \frac{\Delta n}{\pi} \mbox{ atan}(12.63 \frac{ (h-h_{0} )}{\Delta h})$
\end{itemize}

where k is the refractive index gradient with height h.  The
atan()  term above is due to a disturbance located  at  a
height $h_0$ having an extent $\Delta h$ and a refractive strength  $\Delta n$.
For  a  normal atmosphere ($\Delta n=0$ in the Webster model)
 both  models
are linear in h (after expanding the exponential to
first  order).  Nevertheless,  their  dependence  solely  on
height  does not account for the 3D nature of the atmosphere
and  its  disturbances.  Some models  like  the  recent  one
introduced by Costa  \cite{costa} mimics a 3D atmospheric disturbance  by
multiplying the refractive index along the vertical  with  a
Gaussian function along the horizontal perpendicular to  the
ray  path plane. Going beyond these approaches, we introduce
a full 3D profile:

\begin{equation}
 N=p_{x}(x) p_{y}(y) p_{h}(h)+k h+N_{0}
\end{equation}

where $ p_{x},  p_{y}  \mbox{ and }  p_{h}$  are  the  index  profiles  of   the
disturbance along the three directions in space x, y  and  h.
$N_{0}$  is an average normal atmosphere index and k is the index
gradient  along the height. A profile function  $p(X)$,  along
direction X is typically taken as:
\begin{eqnarray}
&& p(X) =(\Delta n_{x}/2)[ tanh((X-X_{1})/\Delta X_{1}) \nonumber \\
&&\hspace{2cm}\mbox{}     -tanh((X-X_{2})/\Delta X_{2}) ]
\end{eqnarray}

where  $X_{1} $ (resp.  $X_{2}$) is the point where  the  hump  starts
growing  (resp. decaying) and $\Delta X_{1}$ (resp. $\Delta X_{2}$) 
is  a  typical length  scale  for  the growth (resp.  decay).  $\Delta n_{x}$  is  the
refractive  strength of the disturbance. This model,  though
realistically    representing   a   localized    anisotropic
disturbance in the atmosphere is based on a separable  model
of the refractive index function.\\

While  our methodology can handle any arbitrary 3D model  of
the refractive index, any of these refractive models have to
be modified in order to take account of the curvature of the
Earth by the inclusion of a term \cite{webster} equal to $10^{6} h/R_{e}$ where $R_{e}$
is the radius of the Earth.

\section{Stiff Differential Equations Algorithms}

Using [4], the ray-tracing system [3] is rewritten as:

\begin{eqnarray}   
   \frac{d {\bf R}}{ds} &=& {\bf T}     \nonumber \\
   \frac{d {\bf T}}{ds} &=& [{\bf grad} N-{\bf T} ({\bf grad}N {\bf .T})]/(N+10^6)
\end{eqnarray}             
          
Two important features appear in the RHS of the second
equation in the system:
\begin{itemize}
 \item The non-linear term in ${\bf T}$.
 \item The wide range of orders of magnitudes in the
denominator.
\end{itemize}

These  terms can be eliminated with the following procedure:
Replace equation [7-b] by another equation defining the curvature of
the ray path $r$:

\begin{equation}
\frac{d {\bf T}}{ds} = {\bf U}/\rho
\end{equation} 

where  ${\bf U}$ is the normal to the trajectory. ${\bf U}$ is perpendicular
to  ${\bf T}$ and  normalized:  $||{\bf U}||=1$.  The  unknown  $\rho$  can  be
determined by taking the scalar product of both sides of 
[7-b] with ${\bf U}$ and using [8]; one gets:

\begin{equation}
 1/\rho={\bf U.grad}N/(N+10^6)
\end{equation} 

Substituting [9] in [8] gives the following system:

\begin{eqnarray}
\frac{d {\bf R}}{ds} &=&  {\bf T}  \nonumber \\
\frac{d {\bf T}}{ds} &=&   {\bf U} ({\bf U.} {\bf grad}N)/(N+10^6)
\end{eqnarray}
 
In general, this system is not closed because it involves ${\bf U}
\mbox{ besides }{\bf R} \mbox{ and } {\bf T}$. In two dimensions, 
one can close the system
by invoking \cite{born} the orthogonality of ${\bf U} \mbox{ and } {\bf T}$ through:

\begin{equation}
 {\bf U} =  {\bf x} \times {\bf T}
\end{equation}

where  ${\bf x}$  is  the  unit  vector along the  x  direction.
 With relation [11], system  [10] is now closed and can be
 integrated by any standard explicit
integration method (Predictor-corrector, Euler, Runge-Kutta,
Richardson etc...). This will be illustrated in section 4.  In
general, $N$ is a function of the position vector {\bf R};  when  it
depends  only  on  the  height, it is  possible  to  further
simplify  the  system  and reduce  it  to  a  single  scalar
equation.  In  the case $N$ depends only on height, ${\bf grad}N$  is
along  the vertical and if $\psi$ is the angle $ {\bf T}$ makes  with  the
local  horizontal, ${\bf U}$ being perpendicular to ${\bf T}$ will make  the
same angle with the vertical, [9] yields:

\begin{equation}
\frac{1}{\rho}= |(dN/dh) cos \psi| /(N+10^6)
\end{equation}

Livingston \cite{living} has derived an equation similar to [12]:

\begin{equation}
\frac{1}{\rho}= -(1/n) (dn/dh) cos \psi                      
\end{equation}
                                                            
Equation [13] is equivalent to [12] when the right  sign  is
used.  We have integrated system [10] in two dimensions  and
recovered typical results found in the literature,  avoiding
the  difficulty arising from [7-b]. In the three dimensional
case,  one  has to deal directly with system  [7]  with  all
terms retained, for, in general, the ${\bf T}$ vector does no longer
have  to  be confined to the transmitter (TX) receiver  (RX)
plane.  In  this  case,  all standard  explicit  integration
schemes break down. In other words, the norm of the vector ${\bf T}$
tangent to the ray path is no longer conserved. In order  to
fulfill   the  condition  $||{\bf R}||=1$,  one  has  to   take   an
integration  step so small that the integration  process  is
virtually   stopped.  This  is  called  stiffness   and   an
illustrative example is given in Appendix A.\\

Stiffness   can  be  cured  with  the  so  called   implicit
integration  schemes.  In contrast to  explicit  integration
schemes  where a current system value depends  only  on  the
previous  ones,  implicit schemes couple  present  and  past
values  of  the  system altogether. A price  to  pay  is  an
increase in CPU time but the rewards are stability, accuracy
and   large  integration  steps.  We  have  implemented  two
implicit schemes:

\begin{itemize}
\item Generalized Runge-Kutta (GRK) method of fourth order \cite{kaps}.
\item Rosenbrock (ROW) method of sixth order \cite{wanner}.
\end{itemize}

In the first scheme, given a system of first order ordinary
differential equations (ODE):

\begin{equation}
d{\bf y}/ds={\bf f(y)}
\end{equation}

one builds the vectors from the system values at step n-1:
\begin{equation}
{\bf k_i}= \sigma {\bf f}({\bf y_{n-1}}+ \sum a_{ij} {\bf k}_{j})    \hspace{2cm}   \mbox{with: i, j=1...m}
\end{equation}

and evaluates the next value n of the system with:

\begin{equation}
 {\bf y_n}= {\bf y_{n-1}} + \sum b_{i} {\bf k}_{i}
\end{equation}                           

$\sigma$ is  the  integration  step  and  the  $a_{ij}  \mbox{ and }  b_{i}$  are
coefficients  depending on the scheme m of  the  integration
order.  In the Rosenbrock case, one adds to [15] the term 
$\sigma (\frac{\partial {\bf f}}{\partial {\bf y}}) \sum d_{ij} {\bf k}_{i}$, 
  where  the  $d_{ij}'$s  are   order   dependent
coefficients and $(\frac{\partial {\bf f}}{\partial {\bf y}})$
 is the Jacobian of the system.  The
above  equations are implicit since the unknown  vectors ${\bf k}_{i}$
needed for integration step n appear on both sides of  [15].
In  the  GRK  method, only the vector function  ${\bf f }$ is  needed
whereas  in  the  ROW  case  both  ${\bf f }$  and  its  first  order
derivative (Jacobian) are needed.\\

Both  methods have been proven to perform very  well  up  to
stiffness  parameters (ratio of the highest to the  smallest
eigenvalue  of  the Jacobian) as high as $10^7$.  Incidentally,
our stiffness parameter has been observed (while testing ROW
algorithms) to be generally around $10^4$. We have used GRK  of
order  4  and  ROW  of  order  6  because  they  have   been
extensively  tested  for a wide range  of  systems  and  are
thoroughly documented.

\section{Validation of the approach and comparisons with previous
treatments}

In  order  to  validate our technique,  we  started  with  a
comparison  against  analytically  known  solutions.   Three
models  were  tested,  the axial gradient  refractive  index
case, the sine-wave optical paths and the classical Luneburg
lens  (see,  for instance, reference 7). In all three  cases
our  results  compared very accurately with  the  analytical
ones. Then we went ahead and proceeded to solve in detail  a
case  well documented in the literature and investigated  by
Webster \cite{webster}  for  various launching angles. This model  is  two
dimensional  (2D)  and  extensively  referred  to   in   the
literature. We use the 2D version of the system of equations
[10]  which is non-linear (N is a non-linear function  of  ${\bf R}$
and a power of ${\bf U}$ appears in [10-b]).

The  integration, started by taking values of {\bf R} and {\bf T }
as the initial location and launching vectors, is done with a first-
order  Euler and fourth order Runge-Kutta methods. The TX-RX
configuration  and propagation conditions are  the  same  as
those given in Table 1 of Webster's \cite{webster} paper.
In  Fig.2 we show the various ray paths between the  TX  and
the  RX for a series of launching angles (taken with respect
to the horizontal) varying from -0.25 up to 0.5 degrees. The
different  launching  angles, we use, are  respectively,  in
degrees: -0.25, -0.20, -0.15, -0.10, -0.05, 0.0, 0.10, 0.20,
0.30,  0.40, 0.50. The refractive index profile used in  the
study is displayed in Fig.3.\\

While  Fig.2  is based on a first order (Euler)  integration
method,  some changes might occur if we rather use a  fourth
order  Runge-Kutta method. In fact, the ray paths  based  on
either  scheme show no appreciable differences  and  compare
well  with the results found earlier by Webster in the  same
conditions. However, some discrepancies appear for  positive
launching  angles  and  are probably due  to  the  different
levels  of  numerical  accuracy between  our  treatment  and
Webster's.  Let  us  recall that in our case  the  numerical
accuracy  is monitored by checking the conservation  of  the
norm  of  ${\bf T}$. In these simulations, it is conserved  with  an
error smaller than $10^{-7}$. In order to compare our results  to
Webster's   directly,  we  derive,  in  the  same   fashion,
recursion  equations for the ray radial  distance  {\bf R}  (taken
from  the center of the Earth) and the angle $\psi$ that {\bf  T}  makes
with the local horizontal.
Referring  to  Appendix  B  and  Fig.4,  we  can  write  the
following relations:

\begin{eqnarray}
R_{2}   & = & R_{1} + ds \mbox{ } sin( \psi_{1} ) \\
\psi_{2}  & = & \psi_{1} + ds \frac {cos(\psi_{1} )}{R_{1}} - sin^{-1}( \frac{ds} { \rho_{1}})
\end{eqnarray}                                     
                              
where the radius of curvature $\rho_{1}$ is given by [12] with
$\psi = \psi_{1}$ and  $dN/dh$  is  taken at the height $R-R_e$ ($R_e$
is  the  Earth radius).  For  a  given  step ds,  one  starts  the  set  of
iterations [17] and [18] with the launching radial  distance
$R_{1}$ and angle $\psi_{1}$. Using the same initial values as before we
retrieve almost the same ray trajectories obtained in Fig.2.
The validity of our results is monitored by the constancy of
the  modulus  of ${\bf T}$ versus 1. Additionally, we  compared  our
results  (Euler  and  Runge-Kutta) to a very  high  accuracy
integration  technique  based on  the  Butcher's \cite{butcher}  algorithm
(seven-stage  sixth-order  Runge-Kutta  scheme).  The  sixth
order  results are virtually identical to the fourth order's
and  Fig.5  depicts  the ray trajectory  obtained  with  the
different levels of accuracy under the same atmospheric  and
launching  conditions. Incidentally, the difference  between
fourth  and  sixth order trajectories in Fig.5  are  on  the
order of a fraction of a millimeter.\\

In   spite  of  the  above  agreement,  which  is  basically
relative, one still has to gauge independently the  accuracy
of  the  results for a selected order and integration  step.
This is done with the following method: Pick an order p  and
an  integration step $\sigma$; integrate once with $\sigma$ and twice with
$\sigma/2$  in  order to reach the same point; define a step  ratio
$\kappa$ from the difference $\Delta$ between the two results:

\begin{equation}
\kappa   = \sqrt[p+1]{2^{p}/(2^{p}-1)(\Delta/ \epsilon)}
\end{equation}
                              
and  monitor the value of k for a given tolerance,  during
integration.  Ideally, we should have $\kappa \le  2$.  In  Fig.6,  we
display  $\kappa$ versus the integration step number for the  first
order (Euler, p=1) case as well as the Runge-Kutta 4-th order
(p=4) and Butcher 6-th order (p=6)
for a tolerance  of 1 millimeter. We use exactly the same
condition as previously and a launching angle of 0.2
degrees. The figure shows clearly the superiority of 4-th and
sixth order methods for the selected step when such a high
accuracy is desired.

\section{Illustrative results and capabilities of the methodology}

We move on to the description of the 3D propagation case and
show, with a simple example, how we evaluate the power  from
the  antenna radiation pattern, the beam spreading  and  the
state of polarization.
We  select a coordinate system such that the TX is somewhere
on  the  z-axis whereas the y-axis is along the TX-RX  line.
The  vertical plane is defined by the z axis and  the  TX-RX
line.   The   beam  spreading  is  evaluated  by   launching
simultaneously several beams in the vertical and  horizontal
planes  with angles differing by a small amount  from  those
characterizing the main beam. The logarithm of the ratio  of
the  surfaces  swept by the different beams at the  receiver
location  gives an estimate of the spreading loss. In  order
to  account  for  the  TX-RX antenna radiation  pattern,  we
simply  recall  that  the  electric  field  radiated  by   a
parabolic  circular aperture antenna at a point  defined  by
its  distance  r  from the main lobe origin  and  making  an
angle $\theta$ with the lobe axis is given by:

\begin{equation}
E(r,\theta)=j \beta E_{0}a[exp(-j \beta r)/r] \mbox{ }J_{1}(\beta a \mbox{ }sin\theta)/\beta  sin\theta
\end{equation}

where  $a $ is  the aperture radius, $E_{0}$ is a reference  field,
$\beta=2\pi /\lambda$  with   $\lambda$  the  wavelength used, 
$J_{1}$  is  the  Bessel function of the first kind and $j= \sqrt{-1} $.
The  antenna pattern is obtained after normalizing the value
of $|E(r,\theta)|$:

\begin{equation}
 f(\theta)=(2/\beta a)|J_{1}(\beta a \mbox{ } sin \theta )/sin \theta|
\end{equation}

Alluding to our choice of axes, if the main lobe is pointing
in  a  direction defined by the angles $\beta, \gamma$ (in the vertical
and  horizontal plane respectively) and we have a ray  along
$\beta', \gamma'$, the angle the ray makes with the main lobe axis is:

\begin{eqnarray}
&& \theta = cos^{-1}(cos \beta \mbox{ } sin \gamma \mbox{ } cos \beta' \mbox{ }sin \gamma ' \nonumber \\
&&\hspace{1cm}\mbox{} + cos \beta \mbox{ }cos \gamma \mbox{ } cos \beta' \mbox{ }cos \gamma'+sin \beta \mbox{ }sin \beta')
\end{eqnarray}         

The power (in dB) is given by $20 \mbox{ } log_{10}f(\theta)$.
The polarization state of a ray rotates, during propagation,
by  an  angle  calculated with the  help  of  the  following
formula:
\begin{equation}
\phi(A,B)= \int_{A}^{B} {ds/\tau}
\end{equation}
                              
where  $A$  and  $B$  represent the two end points  of  the  ray
trajectory;  $\tau$,  the local torsion of the ray  is  different
from  zero when the trajectory is not confined to  a  plane.
Using the Frenet-Serret \cite{born} formula:

\begin{equation}
d {\bf B}/ds=-\tau  {\bf U}
\end{equation}
                               
Taking the dot product with ${\bf U}$ on both sides of equation [24]
and replacing the value of $\tau$ in [23], one gets:

\begin{equation}
\phi(A,B)= - \int_{A}^{B} ds^{2}/ ({\bf dB.U})
\end{equation}    
                              
  In  order to evaluate the polarization rotation of the ray
propagating  from  $A$  to $B$ with [25],  a  finite  difference
approximation ${\bf B}_{n}-{\bf B}_{n-1}$ is used for the differential
 ${\bf dB}$, where
the  subscripts  refer to the integration  step.  The  final
discrete formula for the polarization angle reads:

\begin{equation}
\phi(A,B)=  \sum_{n=1}^{n=N} ds^{2}/ ({\bf B_{n-1}.U_{n}})
\end{equation}
                              
where  $N$  is the number of integration steps between $A$   and
$B$.
For  illustration, we treat two 3D examples.  In  the  first
case,  we  take  a  refractive index model consisting  of  a
refractive layer of finite length along the TX-RX line.  The
linear  extent of the layer is taken respectively as  5,  10
15,  20  and  25  kms. Fig.7 shows the dramatic  effect  the
extent  has  on  the ray path. Incidentally, the  refractive
index  model  along the height is taken as the same  Webster
model  as  before  and  the ray launching  is  made  in  the
vertical plane.
In  the second case, we take a refractive index model  given
by  a  Webster profile along z and a profile $p_{y}(y)$ given  by
[5].  Moreover we take an arbitrary 3D launching  direction.
The  resulting 3D ray trajectory for the selected parameters
listed in the corresponding caption is displayed in Fig.8.

\section{Conclusions and future developments}

We  intend  to use this technique to study the  dynamics  of
microwave  radio signals controlled by unstable  atmospheric
layers.  The instabilities cause short error bursts  lasting
from  many  tens  of micro-seconds to a few  milliseconds
 \cite {nigrin91}.
Since,   the  error  bursts  have  detrimental   impact   on
communication  networks \cite {nigrin93}, the future digital radio  systems
should  be  made  immune  to radio propagation  degradations
causing them. In order to develop defense strategies against
the   error   bursts   caused  by  atmospheric   propagation
instabilities,   the   physical   characteristics   of   the
instabilities  have  to  be well understood.  This  3D  ray-
tracing  technique  will be used to  study  the  effects  of
dynamically changing atmospheric layers of limited  size  on
microwave  radio signals received simultaneously  by  a  few
parabolic  antennas \cite {tannous}. A propagation  model  simulating  the
recorded  dynamics  of received radio  signals \cite {nigrin91}  will,  not
only,  help understanding the physical causes of  the  error
bursts,   but   it  will  also  be  used  in  the   computer
optimization  of antenna designs capable of  minimizing  the
frequency  of  occurrence  of the propagation  caused  error
bursts.  Highly accurate numerical techniques  are  required
since  small fluctuations of the atmospheric conditions  are
believed  to  be responsible for the flat phase fluctuations
impairing the digital demodulation of the received microwave
radio signals.\\

\vspace{1cm}

\centerline{\Large\bf APPENDIX A}
\vspace{1cm}
                              
Let us consider the following first order system consisting
of a pair of linear ordinary differential equations:

\begin{eqnarray}
  dy_{1}/dx  & = & \lambda_{+}y_{1} + \lambda_{-}y_{2}  \\
  dy_{2}/dx  & = & \lambda_{-}y_{1} + \lambda_{+}y_{2}
\end{eqnarray}
                              
where  $ \lambda_{+} = (\lambda_{1}+\lambda_{2})/2, \mbox{ }  \lambda_{-} = (\lambda_{1}-\lambda_{2})/2 \mbox { and } x \ge 0$.\\

The solution of the system is:

\begin{eqnarray}
y_{1} & =  &  C_{1} exp(\lambda_{1}  x)+ C_{2} exp(\lambda_{2} x)  \\
y_{2}  & =  & C_{1} exp(\lambda_{1} x)-C_{2} exp(\lambda_{2} x)
\end{eqnarray}
                              
where  $C_{1}$  and $C_{2}$ are constants determined by  the  initial
condition at x=0.
In  order to conform to our notation of Section 3, we define
a  column vector y whose components are $y_{1}, y_{2}$ and write the
system as:

\begin{equation}
d{\bf y}/dt= {\bf f(y)}
\end{equation}
                     
The eigenvalues of the Jacobian of the system:

\begin{equation}
\frac{\partial {\bf f}}{\partial {\bf y}}= \left[ \begin{array}{c c}
\lambda_{+} \lambda_{-} \\
\lambda_{-} \lambda_{+} \end{array} \right]
\end{equation}

are solutions of:
\begin{equation}
det | \lambda  {\bf I} - (\frac{\partial {\bf f}}{\partial {\bf y}}) |=0
\end{equation}

where ${\bf I}$ is the (2x2) unit matrix; they are nothing else than
$\lambda_{1}$  and  $\lambda_{2}$.  If one picks 
$\lambda_{1}=-1, \lambda_{2}=-1000$. and  chooses  an
explicit integration method, one finds the integration  step
should be smaller than $2/|\lambda_{2}|$ , which is 0.002. This is  the
origin  of  stiffness: even though the  term  exp(-1000 x)
contributes  almost  nothing to the solution  for  $x \ge 0$,  its
presence alone, virtually stops the integration process.\\

\vspace{1cm}

\centerline{\Large\bf APPENDIX B}

\vspace{1cm}
                              
The  geometry of propagation is shown in Fig.4. At any point
along  the  ray trajectory the tangent vector $ {\bf T}$  makes  the
angle $ \psi$  with the local horizontal. When the ray propagates
between two nearby locations, one may write:

\begin{eqnarray}
R_{2}   & = & R_{1} + ds \mbox{ }sin( \psi_{1} ) \\
\psi_{2}  & = & \psi_{1} + ds \frac {cos(\psi_{1} )}{R_{1}} - sin^{-1}( \frac{ds} { \rho_{1}})
\end{eqnarray}

\begin{equation}
R_{2}   =  R_{1} + ds \mbox{ } sin( \psi_{1} )
\end{equation}
                              
where $ ds=|| {\bf R}_{2} - {\bf R}_{1}||$ . The radial distance $R_{1}$ (resp. $R_{2}$) is
taken from the center of the Earth. The angle $\delta \theta$ between the
two radial directions may be found by inspection:

\begin{equation}
R_{1} sin(\delta \theta)= ds \mbox{ } cos( \psi_{1})
\end{equation}
                              
which can be approximated by:

\begin{equation}
\delta \theta=ds \mbox{ } cos( \psi_{1})/R_{1}
\end{equation} 

In order to find the relation between the angles $\psi_{1} \mbox{ and }  \psi_{2}$,
we use the relation defining the derivative of ${\bf T}, d{\bf T}/ds={\bf U}/\rho$ in a discrete form:

\begin{equation}
  {\bf T}_{2} - {\bf T}_{1} = ds \mbox{ } {\bf U}_{1}/ \rho_{1}
\end{equation}

Taking the scalar product with ${\bf U}_{1}$ on both sides of above, one
gets:
\begin{equation}
    {\bf T}_{2}{\bf .U}_{1}= ds/ \rho_{1}
\end{equation}

The inspection of Fig.4 provides the angle between $ {\bf T}_{2} \mbox{ and }{\bf U}_{1}$:

\begin{equation}
        ({\bf T}_{2}, {\bf U}_{1})= \psi_{2}- \psi_{1} - \delta \theta +\pi/2
\end{equation}

Using the above result gives the relation sought:
\begin{equation}
       \psi_{2}=  \psi_{1}+ \delta \theta  - sin^{-1}(ds/ \rho_{1})
\end{equation}

\vspace{1cm}
\centerline{\Large\bf Figure Captions}
\vspace{1cm}

\begin{itemize}

\item[Fig.\ 1:]  Geometry  of the system showing  the  coordinate
system, the antennas in the vertical yz plane, a typical ray
path  and  the local Frenet-Serret system $({\bf T, U, B})$ attached
to a point along the path.

\item[Fig.\ 2:]  Euler  first  order 2D  results.  The  rays  are
launched in the vertical plane and the angle they make  with
respect to the horizontal xy plane is respectively: -0.25, -
0.20, -0.15, -0.10, -0.05, 0.0, 0.10, 0.20, 0.30, 0.40, 0.50
degrees. Equations [10] are used along with model [4-b]  for
a  perturbed atmosphere $N=300.+kh+ \frac{\Delta n}{\pi} \mbox{ atan}(12.63 \frac{ (h-h_{0} )}{\Delta h})$
with  the  same  parameters as those given  in  Table  1  of
reference 2: $k=-39, \Delta n=-20$ (both in N units), $h_{0}$=175 meters,
$\Delta h$ =100 meters, the transmitter height is 125 meters and  the
TX-RX separation is 60 kms.

\item[Fig.\ 3:] Webster \cite{webster} model refractive index function  (in  N
units) along the vertical showing an anomaly at a height  of
175  meters  and  whose width is equal to  100  meters.  The
curvature  of  the  Earth  term $10^{6}  h/R_{e}$  is  present.  The
negative   gradient  of  the  layer  refractive   index   is
responsible for the multipath effects observed.\\

\item[Fig.\ 4:]  Geometry  of  the  ray  trajectory   used   for
establishing  the recursion equations. The local  tangent  ${\bf T}$
vectors  are  shown  making  the  angle  $\psi $  with  the  local
horizontal perpendicular to the ray vectors R drawn from the
center of the Earth O. Two neighboring points along the  ray
paths are shown.\\

\item[Fig.\ 5:] Comparative study of the ray trajectories obtained
from  the recursion relations [17] and [18] (uppermost  long
dashed  curve) and 1st order Euler (full line curve) on  one
hand,  and  the  Runge-Kutta (4-th order)  and  Butcher  (6-th
order)  on  the  other (short dashed curve).  The  launching
angle in all cases is 0.2 degrees in the vertical plane  and
the model considered for the refracting layer is the same as
Figure  2.  The fourth and sixth order results are virtually
identical.\\

\item[Fig.\ 6:] Comparative study of the behavior  of  the  step
ratio  versus  step number for the Euler  (1st  order),  the
Runge-Kutta (4-th order) and the Butcher (6-th order)  methods
when  the step is fixed to its starting value. Ideally, this
ratio  should  always be  about 2. In the first order  case,  the
bound is violated very rapidly (upper curve), whereas it  is
respected until almost the end of the trajectory in the  4-th
(long  dashed  curve)  and 6-th order  (short  dashed  curve)
cases.  The  tolerance  is 1 mm and  the  step  used  is  one
hundredth the TX-RX distance.\\

\item[Fig.\ 7:] GRK (Implicit, 4-th order, 3D) results for the  ray
trajectories  when the extent of the layer  is  a  variable.
Starting  with  a  launching angle of  0.2  degrees  in  the
vertical  plane, the layer spans, initially, the entire  hop
of  60 kms (lowest curve). Moving upward from the next lower
curve, the layer extent (along the TX-RX line) is from 5  to
25  kms, then 5 to 20 kms, 5 to 15 kms and finally 5  to  10
kms. In all cases, the refracting layer model is the same as
in Figure 2.

\item[Fig.\ 8:] GRK (Implicit, 4-th order, 3D) results for the  ray
trajectory  when  the refractive index of the  layer  varies
along  two  spatial  directions (y and z)  and  round  Earth
profile  considered.  The normal atmosphere  parameters  are
$N_{0}$=300  N  units and the gradient k=-39 N units/km.  The  3D
refractive index layer is described with a profile  along  y
given by $ [ tanh((y-y_{1})/ \Delta y)-tanh((y-y_{2})/\Delta y) ]/2$ with $y_{1}$= 0  km,
$y_{2}$=  60.kms,  $\Delta y$ =100 meters and a Webster  profile  along  z
given  by  $ \Delta n \mbox{ } atan[12.63 (z-h_{0})/\Delta h]/ \pi$ with 
$h_{0}$ =  175 meters,
$\Delta h$ =100  meters  and $\Delta n$= -20.0 N units. The launching  angles
are 0.1, 0.2, 0.3, 0.4 and 0.5 degrees in the vertical plane
with 0 and 0.001 degrees in the horizontal plane. The TX  is
at  125  meters  along  the z axis  and  the  TX-RX  antenna
separation is 60 kms. \\

\end{itemize}

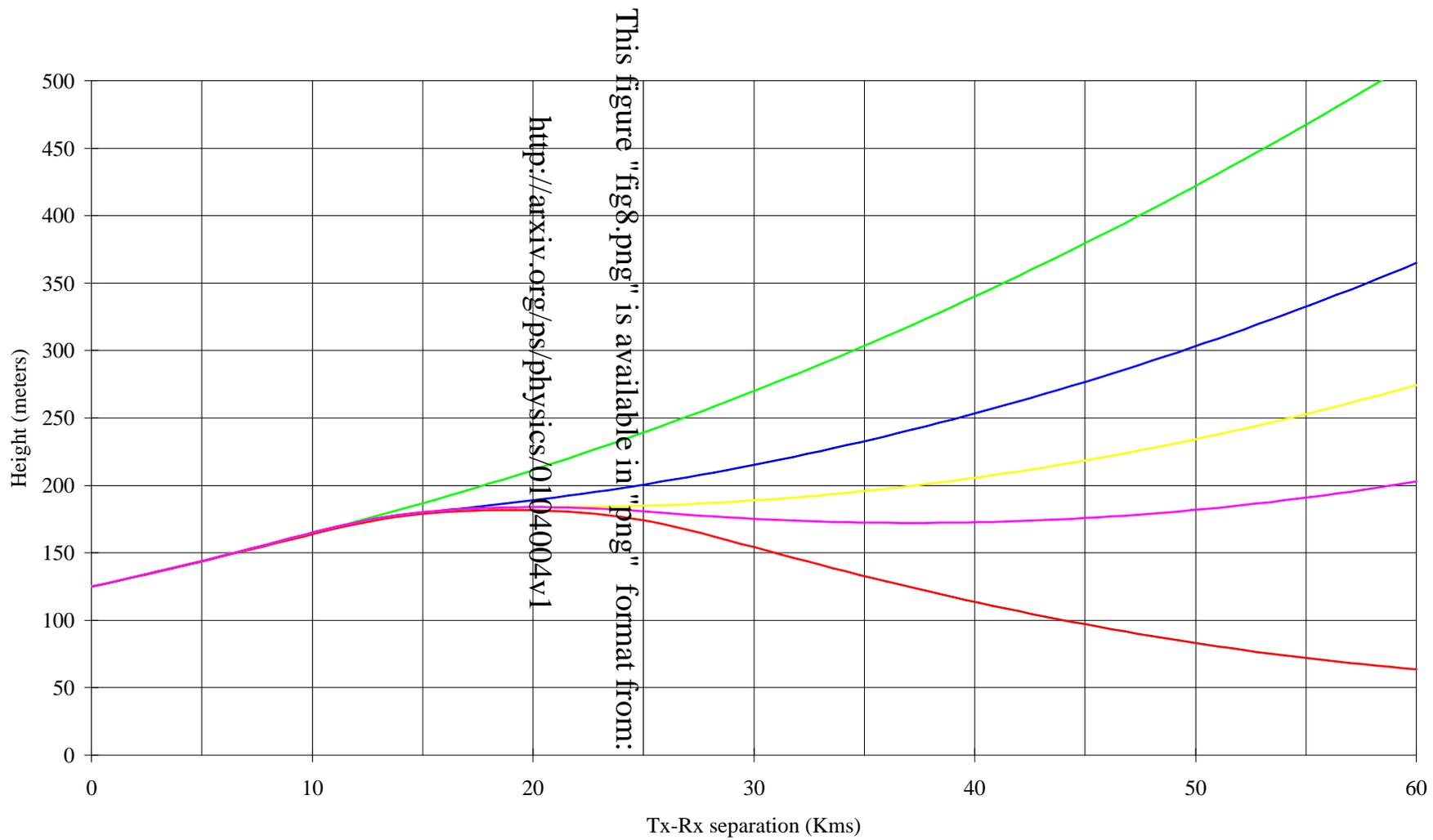
\end{document}